\documentclass[epjST]{svjour}
\usepackage{graphics}
\usepackage[hyphens]{url}
\usepackage[breaklinks]{hyperref}
\hypersetup{
colorlinks=true,
citecolor=blue,
linkcolor=blue,
filecolor=blue,
urlcolor=blue
}

\begin{document}
\title{Accelerating Scientific Discovery by Formulating Grand Scientific Challenges}
%\subtitle{subtitle}
\author{Dirk Helbing\thanks{\email{dhelbing@ethz.ch}} }
\institute{ETH Zurich, Swiss Federal Institute of Technology, Clausiusstr. 50, 8092 Zurich, Switzerland}
\abstract{
One important question for science and society is how to best promote scientific progress. Inspired by the great success of Hilbert'€™s famous set of problems, the FuturICT project tries to stimulate and focus the efforts of many scientists by formulating Grand Challenges, i.e. a set of fundamental, relevant and hardly solvable scientific questions.
} %end of abstract
\maketitle
\section*{Introduction}
\label{intro}

When David Hilbert formulated Grand Challenges in mathematics about a century ago [1], this strongly promoted the development of modern mathematics, even though some findings were quite different from what was expected (the famous incompleteness theorem by Kurt G\"odel demonstrates this well). To stimulate scientific progress, it has therefore been proposed to formulate `€˜Hilbert problems'€™ also in other research areas [2-5].
Of course, many Grand Challenges in social and economic sciences will not have an exact mathematical answer. Moreover, some problems may not have a solution at all, but it could still be possible to reach significant improvements. Despite the considerable differences between mathematical and real-world problems, a crucial part of performing science is to identify good questions. Without this foundational step, scientific discovery is likely to be erratic. Good questions can guide the way towards new discoveries. Once a question is formulated, it is often just a matter of time, until progress is made.
\par
To stimulate this progress, outstanding scientific publications addressing challenges such as the ones listed below should be rewarded by prestigious research prizes, stipends, or grants, even when the answers are only steps forward rather than a complete solution. These rewards should be distributed by a high-level international, multi-disciplinary jury (science board) according to the progress made. But it is even more important that the academic system also provides incentives for identifying good questions and sharing them with others. Currently, the publication of open research challenges or a list of such challenges is rarely found in conventional peer-reviewed journals. This must be changed.
Therefore, over the time period of its implementation, the FuturICT project [6] intends to create an open platform, where Grand Challenges can be publicly posted by scientists, citizens or institutions. FuturICT'€™s envisaged crowd funding platform shall help to match questions, funding and ideas. Sponsors (such as companies, individuals, funding agencies, or non-profit organizations) could then provide a budget for the solution of these questions. Such problems, when attached with money, will attract the interest of bright minds. The financial rewards would allow them to get money for future research.
\par
According to this new funding principle, money will be provided for the best solutions, not for proposals or promises. A positive side effect would be that the level of proposal writing and reporting could be considerably reduced, which currently impairs scientific productivity a lot. This novel approach of organizing and supporting science could establish a new research and innovation paradigm (see Refs. [2,7] for a more detailed discussion).
\par
We now turn to the central theme of this paper, namely the formulation of a set of challenging scientific questions (see also Refs. [3-5]). Some of the problems have certainly been around for quite a while, but still require attention and a federated effort.
\par
\section*{I. Socio-Economic Real-World Challenges}
\begin{enumerate}
  \item How to reach a balance of power in a multi-polar world (between different countries and economic centers, between the worlds of business and politics, between individual and collective rights)?
\item	How to promote security and peace (e.g. avoid organized crime, terrorism, social unrest)?
 \item What are the contributing factors and dynamics of conflict? How to avoid, overcome or moderate conflict, or turn it into a creative force? How to facilitate a peaceful interaction of people with incompatible values and diverse cultural backgrounds?
 \item 	What contributes to the spreading of crime and corruption, and how to counteract it?
 \item	What is the origin of social and economic inequality? How can poverty and precarious living conditions be reduced? How much inequality is beneficial for socio-economic progress, and how can it be stabilized?
 \item	How to increase the quality of life, satisfaction, and well-being of people? How to reduce suicide rates?
 \item	How to promote public health (increase food safety; reduce the spreading of epidemics, obesity, smoking, or unhealthy diets...)?
\end{enumerate}

\section*{II. Measurement and Methodological Issues}
\begin{enumerate}
\item	How to measure characteristics of society, the social fabric, social norms, social capital, social impact, social change, interactions, systemic risks, institutional constraints, context, and culture globally and in real-time?
\item	How to operationalize and verify or falsify qualitative interpretations in the social sciences? How to do qualitative social analyses with computers? How to translate qualitative social knowledge into social mechanisms that can be simulated in an agent-based way?
\item	How to bridge between qualitative and quantitative social science approaches, e.g. overcome the gap between abstract concepts and concrete contents? To what extent can cultural developments be understood as a result of interactions, history and context? What roles do networks of people and institutions, co-location effects, and randomness play?
\item	How to perform large-scale behavioral experiments in an efficient and ethically acceptable way? How to integrate experimental data and theories of the social, economic, behavioral, cognitive, and neuro-sciences into computational social science?
\end{enumerate}

\section*{III. Understanding Underlying Principles}
\begin{enumerate}
\item Are there explanatory principles underpinning human societies and, if yes, which ones? What principles allow socially interactive systems to work well? What are the driving forces underlying the dynamics and evolution of these systems?
\item	How to develop accurate, predictive and quantitative models that link individual action to social phenomena and policy change to individual action?
\item	How to model real (rather than idealized) economic systems? What contributions can models of ecological systems make to the understanding of complex techno-socio-economic systems?
\item	Under what conditions do concepts such as Adam Smith's `€˜invisible hand'€™, equilibrium, efficient markets, or rational agents work well? What are their limitations?
\end{enumerate}

\section*{IV. Modeling of Complex Adaptive Agents}
\begin{enumerate}
\item	How do individuals use available information resources to take decisions?  Is bounded rationality an important factor in individual decisions and, if yes, under what conditions? How does this affect emergent macro phenomena arising from interactions of many agents? When do representative agent models fit observed phenomena well?
\item	How to model cognitive complexity, subjectivity, emotions, and learning? How to realistically and efficiently model individuals as computer agents with cognition? How to model the interaction of emotional states with belief-based processes? What are the roles of individual context, social learning, and self-image for individual behavior?
\item	What level of complexity of computer agents is needed to get a good representation of the dynamics and outcomes of social interactions? In particular, how much complexity is needed to model agents that form social habits and routines and are nonetheless able to overcome them to gain full control over their behavior, when required by the context?
\end{enumerate}

\section*{V. Interactions, Networks, and Systemic Interdependencies}
\begin{enumerate}
\item	What phenomena may occur in strongly coupled and interdependent systems and under what conditions? How to characterize these phenomena? What universality classes exist?
\item	What are the interdependencies between structure, dynamics and function? How do networks affect the behavior of their components? Are there properties that individual components 'inherit' from the networks they belong to?
\item	What are the limits of predictability and controllability in complex, networked systems? Is the economic system designed in a controllable€™ way (from a cybernetic point of view)? If not, what would have to be changed to make the system better manageable?
\item	What institutional designs and interaction rules can create resilient functional networks in a decentralized way? How effective is this?
\end{enumerate}

\section*{VI. Dynamics and Change}
\begin{enumerate}
\item	How to anticipate the implications of demographic change (e.g. population growth, aging, migration)? For example, what is the impact of demographic change on settlements? How to cope with demographic change better? How to create adaptable cities? How to create social benefit systems that can adjust to demographic change? How to influence demographic developments?
\item	How do environment and environmental change interact with human behavior and social change?
\item What are the effects of (current and future) information and communication technologies on societies? How do data-driven and model-enabled global information systems change the decision-making of people and the structure of organizations? Will personalized services and recommendations promote consensus, individualism, or conflict? What role does empathy play in our information society?
\item	What (institutional) relationships between the state, the law, private enterprises and individuals can ensure and enhance human rights and social well-being in an increasingly data-rich world? How can the regulatory and legal system deal with emergent technologies? What are legal challenges and limitations of future information and communication systems? How to enable an information ecosystem that co-evolves with society?
\end{enumerate}

\section*{VII. Self-Organization, Collective Behavior, Co-Evolution, and Emergence}
\begin{enumerate}
\item	When do correlations matter and generate a co-evolutionary dynamics that cannot be captured by a mean-field approximation or `representative agent approach'? Which network structures promote aggregate effects through correlations, which ones average over neighboring interactions?
\item	When and how do heterogeneity (diversity) and randomness (fluctuations, noise) change the system dynamics fundamentally?
\item	How to understand collective social behavior and opinion dynamics (e.g. abrupt changes in consumer behavior; social contagion, extremism, hooliganism, changing values; breakdown of cooperation, trust, compliance, solidarity...)? What are the effects of social contagion on the diffusion of innovations, both on the epistemic and behavioral levels?
\item	What are the factors determining the speed and range of the spreading of different opinions, beliefs, ideologies, and new ideas etc.? How does consensus emerge, and to what extent can it be predicted? When do self-fulfilling prophecies occur, when self-defeating ones, and when does the system behave in an inert way?
\item	Under what conditions do quantitative changes make a qualitative difference (`€˜more is different'€™) or vice versa? How do qualitative and quantitative features affect each other?
\end{enumerate}

\section*{VIII. Risks, Systemic Instability, and Resilience}
\begin{enumerate}
\item	Under what conditions is the dynamics of a system sensitive to probabilistic events ('noise'€™)? How do extreme events come about? To what extent do network interdependencies create new forms of (potentially catastrophic) risks? How to quantify, measure, and communicate systemic risks?
\item	How to understand systemic shifts (e.g. from peace to war, from autocratic to democratic systems, from individual to collective behavior)?
\item	How to cope with systemic risks and extreme events? In a more connected world, can we design systems in a way that avoids large-scale risks related to contagious spreading and cascade failures?
\item	How to create resilient networks and systems? How to avoid that local perturbations can have destructive systemic impacts on a global scale?
\end{enumerate}

\section*{IX. Markets, Transaction and Exchange Systems}
\begin{enumerate}
\item	How to avoid or cope with financial and economic (in)stabilities (explosion of government debts; inflation; overloaded social benefit systems; abrupt variations in consumption and investment behavior...)?
\item	How to increase the stability of the global financial system? What set of properties should a financial system have in order to meet societal needs and what financial architectures would promote them? What are the properties of alternative banking, pricing, auctioning, market and exchange systems (like efficiency, predictability, reliability, robustness, signs of failure, etc.)? How to create financial markets without bubbles and crashes? How to reduce already existing bubbles? How to solve the too big to fail problem?
\item	How to manage inefficient and imperfect markets? How to reduce the manipulation of market prices (e.g. through artificial shortage)?
\item	What is the impact of financial innovations on the stability of the economic system and what does it depend on? Can low-latency/high-frequency trading destabilize financial markets? What monetary policies can stop financial contagion and overcome financial crises?
\item	How to establish a more efficient and robust exchange of value(s) between agents, considering mechanisms like trust, reputation, and norms? How to create a peer-to-peer (P2P) financial system for payment and lending? How to overcome the limitations of one-dimensional money and create multi-dimensional incentive systems?
\item	How to design auctioning mechanisms for energy markets that can handle decentralized energy production without large price fluctuations and cannot systematically be exploited (i.e. which are fair to everyone)?
\item	How to balance economic and social conflicts of interest?
\end{enumerate}

\section*{X. Institutions and Integrative Systems Design}
\begin{enumerate}
\item	In what ways do decision-making rules determine the outcome and behavior of social systems? How to organize decision-making and delegate power in a way that maximizes the satisfaction of people? How to take wise collective decisions that avoid herding effects, overconfidence, and a suppression of minorities?
\item	How do structures, hierarchies, and institutions emerge and what are the properties associated with them? How to design social, economic, and political institutions in a way that allows us to address social problems more successfully? How to understand the interaction of people with institutions and the interaction between institutions?
\item	How do bottom-up and top-down processes of social regulation interact? How to reach the best combination of top-down (centralized, global) and bottom-up (decentralized, local) decisions?
\item	How to develop mechanisms and institutional settings (such as regulatory and legal frameworks), which establish resilient social, economic and political systems that seamlessly adapt to global change? How to avoid over-regulation and corruption, etc.?
\item	How to increase opportunities for social, economic, and political participation (of people of different gender, age, health, education, income, religion, culture, language, preferences)? How to minimize unemployment? How to reduce the delegation of decisions, where people have their own, separate interests?
\end{enumerate}

\section*{XI. Management of Complexity and Sustainability}
\begin{enumerate}
\item	How to manage complexity and support favorable self-organization? How to simulate the impact of policy-making on society?
\item	What is the interplay between economic growth, social capital, and social well-being? How to create a sustainable economic system that performs well without a steady need to grow? Can one maintain (or even increase) human well-being without economic growth and, if so, how? What are the social, political, and technological requirements for sustainable and resilient economies?
\item	What incentive systems can support efficient coordination and overcome social dilemmas (`€˜tragedies of the commons'€™)? How to promote a sustainable use of environmental, social, economic, and technological resources? How to support favorable consumption habits, travel behavior, sustainable and efficient use of energy, participation in recycling efforts, environmental protection?
\end{enumerate}

\section*{XII. Information, Innovation, and Knowledge}
\begin{enumerate}
\item	How to aggregate information in a transparent and accountable way? How is information transformed into knowledge and what decides about the societal impact of different kinds of information? What determines the interpretation, meaning and impact of certain kinds of information? How does information influence the behavior of people? How do complex systems aggregate and respond to information?
\item	How does innovation arise? What drives or impedes innovation? How to model creativity?
\item	How to do information management in an age of Big Data (considering issues such as possible cyber risks, espionage, violation of privacy; data deluge, spam; education and inheritance of culture...)?
\item	Under what conditions is data mining anonymity-preserving, even when combined with other sources of information? What characterizes sensitive data and what to do about them, how to protect them, and avoid misuse? How to best reduce sensitive information in personal or commercial datasets?
\item	How to organize intellectual property rights in a way that benefits innovators, individuals, and society better?
\end{enumerate}

\section*{XIII. Social and Economic Capital, and New Forms of Value}
\begin{enumerate}
\item	How to distribute scarce resources in a way that supplies those in need even if not financially well equipped? How to avoid or mitigate global shortages? How to promote fair trade and exchange?
\item	How can future ICT systems support social innovation? How can they help to protect and create currently existing and new forms of human, social, and economic capital? How to model, measure, and understand the emergence and breakdown of social capital? How to understand and model social and economic value as an emergent property?
\item	How to promote the generation of new intangible goods (and markets for them) with future information and communication systems? How to create a self-regulating trustable web? How to design a decentralized multi-dimensional rating, manipulation-resistant reputation, and pluralistic recommender system? How to design recommender systems that maintain a healthy degree of socio-diversity? How to measure and reward incremental collective inventions and innovations? How to share intellectual property in a crowd sourcing process?
\item	What determines the legitimacy of rules and procedures, and the compliance with them (i.e. with laws, regulations, etc.)?
\end{enumerate}

\section*{XIV. Values, Responsibility and Ethics}
\begin{enumerate}
\item	How to build a high-quality data commons? How to keep harmful applications of information and communication systems on a low level and avoid undesirable systemic impacts? How to promote a responsible use of exchange, information, and communication systems? How to develop ethical information and communication systems and value-sensitive designs?
\item	How to create accountability and responsibility, if responsibility is distributed among many people? How to promote responsible behavior and awareness for the impact of human decisions and actions on our techno-socio-economic-environmental system?
\item	How to support the application of ethical principles in a world of profit maximization, without enforcing them by law? How to run an economy, in which the compliance with ethical values does not create a competitive disadvantage?
\item	How to create conditions in which the public and private sphere benefit each other better? How to perform data mining in a way that can serve public and commercial interests, but is compatible with individual interests (such as privacy)?
\end{enumerate}

\section*{Conclusions}

Of course, one cannot expect that the FuturICT project can solve the above questions over a 10 years time period. However, one can hope that the above list of Grand Challenges will stimulate the research of many scientists in different disciplines all over the world and that this will accelerate the generation of knowledge on important problems, which deserve more attention and support.
Note that a more precise specification of the above problems is part of the scientific challenge, because the solution is often not a mathematical proof, but rather a matter of empirical or experimental evidence, and may depend on the value system or priorities. If researchers develop different perspective on the same problem, this can be very fruitful.

\subsection*{Acknowledgments}
The author would like to thank Stuart Anderson, Anna Carbone, Rosaria Conte, and an anonymous referee for their valuable comments and suggestions, and many others for interesting discussions.

The publication of this work was partially supported by the European
Union's Seventh Framework Programme (FP7/2007-2013) under grant agreement no.
284709, a Coordination and Support Action in the Information
and Communication Technologies activity area (`FuturICT' FET Flagship Pilot Project).

\end{document}